\documentclass[a4paper]{jpconf}
\usepackage{graphicx}
\usepackage{amsmath}
\usepackage{amssymb}
\usepackage{ulem}
\usepackage{mathrsfs, mathtools, mhsetup,slashed}
\usepackage{epstopdf}
\usepackage{bm}% bold math

\newcommand{\ourprocess}{ p \bar{p} \,\rightarrow\, \overline{D^0} D^0}

\begin{document}

 \title{Production of heavy meson pairs in $p\bar{p}$ collisions within a double handbag approach}
 \author{A T Goritschnig$^{1a}$, B Pire$^{2b}$ and W Schweiger$^{1c}$}
 
 \address{$^1$ Institute of Physics, University of Graz, Universit\"atsplatz \( 5 \), \( 8010 \) Graz, Austria} 
 \address{$^2$ Centre de Physique Th\'{e}orique, \'{E}cole Polytechnique, CNRS, 91128 Palaiseau cedex, France} 
 
 \ead{$^a$alexander.goritschnig@uni-graz.at}
 \ead{$^b$bernard.pire@cpht.polytechnique.fr}
 \ead{$^c$wolfgang.schweiger@uni-graz.at}

\begin{abstract}
We study the pair-production of heavy mesons in proton-antiproton annihilations within a perturbative QCD-motivated framework.  
In particular we investigate $p \bar{p} \rightarrow \overline{D^0} D^0$ within a double handbag approach, 
where a hard subprocess factorizes from soft hadronic matrix elements. 
The soft matrix elements can be parametrized by transition distribution amplitudes, which are off-diagonal in flavor space. 
The transition distribution amplitudes are modelled as overlaps of light-cone wave functions. 
We obtain rather robust model results for $p \bar{p} \rightarrow \overline{D^0} D^0$ cross sections, 
which are expected to be measured at the future $\bar{\text{P}}\text{ANDA}$ detector at GSI-FAIR.
\end{abstract}

\section{Introduction}
\label{sec:Introduction}
 
In Ref.~\cite{Goritschnig:2012vs} we have studied the process $\ourprocess$ which has become of particular interest 
due to the planned physics program of the $\bar{\text{P}}\text{ANDA}$ detector \cite{Lutz:2009ff} at FAIR-GSI. 
A part of the FAIR project with its $\bar{\text{P}}\text{ANDA}$ detector is devoted to the measurement of 
exclusive channels in proton-antiproton collisions where heavy hadron pairs are produced. Therefore theoretical input is most welcome. 
To investigate $\ourprocess$ we use the framework of perturbative QCD where we apply a generalization of the, so-called, \lq\lq handbag approach\rq\rq. 
A handbag mechanism has been first developed to study deeply virtual Compton scattering and meson production \cite{mueller94}; 
in Ref.~\cite{Goritschnig:2009sq} it has been extended to the flavor off-diagonal case, which comes across here. 
For our studies of $\ourprocess$ we go along similar lines of argumentation as for 
$p\bar{p} \,\rightarrow\, \Lambda_c^+ \overline{\Lambda}_c^-$ in Ref.~\cite{Goritschnig:2009sq}. 
Under physically plausible assumptions we argue that our process under consideration factorizes into a 
hard partonic subprocess amplitude and soft hadronic transition matrix elements. 
In doing so the hard energy scale justifying this splitting and the treatment of 
the partonic subprocess within perturbation theory is set by the heavy charm-quark mass $m_c$. 
In our investigations we only take the valence Fock states of the hadrons into account and 
consider the proton in a quark-diquark picture. 
The soft hadronic transition matrix elements can in principle be parametrized by transition distribution amplitudes \cite{PSS}. 
We here represent them as an overlap of hadron light-cone wave functions (LCWFs) analogously as in Ref.~\cite{Diehl:2000xz}.

\section{The Double Handbag Mechanism}
\label{sec:The-Double-Handbag-Mechanism}

\begin{figure*}
  \includegraphics[width=0.48\textwidth, height=0.25\textwidth]{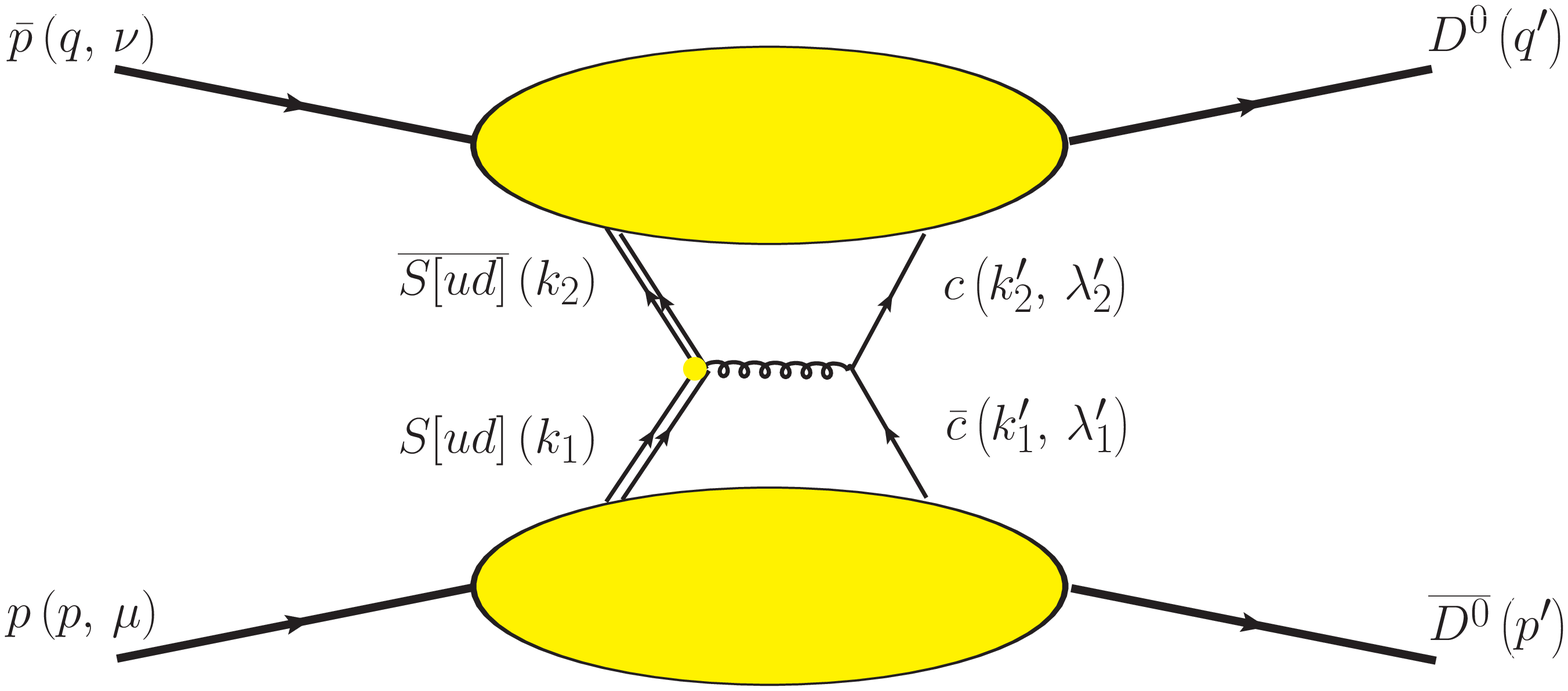}\hfill
  \includegraphics[width=0.48\textwidth, height=0.25\textwidth]{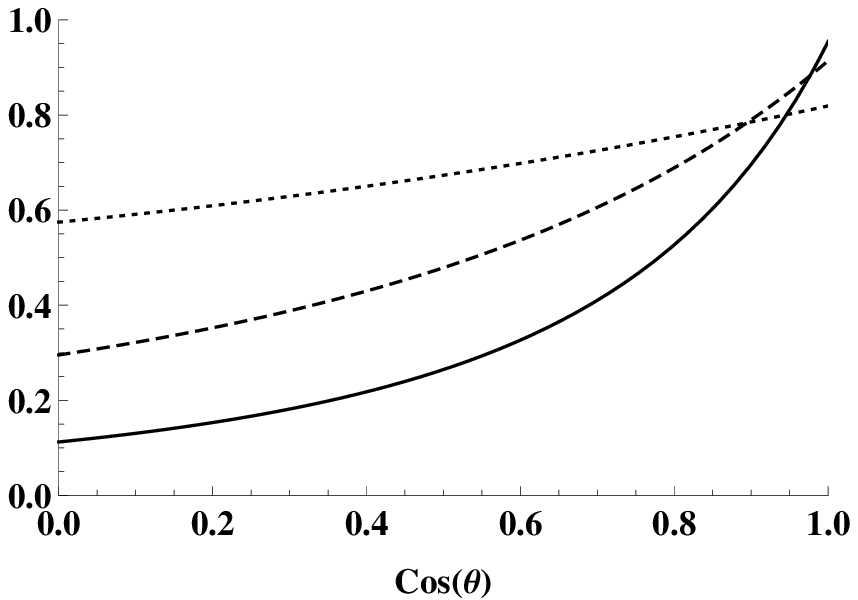}
\caption{{\it Left:} The double handbag mechanism for the $\ourprocess$ scattering amplitude. 
Blobs represent the soft $p \,\to\, \overline{D^0}$ and $\bar{p} \,\to\, D^0$ contribustions. 
{\it Right:} The $\bar{x}$-moment of the wave function overlap appearing in the  
square brackets in Eq.~(\ref{eq:DHM-amplitude-peaking-approximation})
versus $\cos\theta$ for $s=15,\,20,\,30$~GeV$^2$ (dotted, dashed, solid).}
\label{fig:overlap}
\end{figure*}

Before we turn to the description of $\ourprocess$ within a double-handbag approach 
we would like to specify the kinematics chosen to describe the process. 
The assignment of particle momenta and helicities can be seen in Fig.~\ref{fig:overlap}.
$M$ and $m$ denote the masses of the $\overline{D^0}$ and the proton, respectively. 
We work in a center-of-momentum system (CMS) where the $z$ axis is parallel to the 3-vector 
$\mathbf{\bar{p}}$, with $\bar{p} := (1/2) (p^\prime + p)$, and the transverse momentum transfer ${\bm \Delta}_\perp := (\Delta^1,\,\Delta^2)$, 
with $\Delta := p^\prime - p$, is symmetrically shared between the incoming and outgoing hadron momenta. 
In light-cone coordinates\footnote{A four vector $v=(v^0,\,v^1,\,v^2,\,v^3)$ written in light-cone coordinates reads $v = [v^+,\,v^-,\,v^1,\,v^2]$ 
where $v^\pm := \frac{1}{\sqrt{2}} (v^0 \pm v^3)$.}, which are ideally suited for the description of scattering processes at high energies,  
the proton and the $\overline{D^0}$ momentum read: 
\begin{equation}
\begin{split}
p \,=\,&
\left[ (1+\xi)\bar{p}^+,\, \frac{m^2+\boldsymbol{\Delta}_\perp^2/4}{2(1+\xi)\bar{p}^+},\, -\frac{{\bm \Delta}_\perp}{2} \right] \,,\quad
p^\prime \,=\,
\left[ (1-\xi)\bar{p}^+,\, \frac{M^2+\boldsymbol{\Delta}_\perp^2/4}{2(1-\xi)\bar{p}^+},\, +\frac{{\bm \Delta}_\perp}{2} \right] \,.
\label{eq:particle-momenta}
\end{split}
\end{equation}
The antiproton momentum $q$ and the $D^0$ momentum $q^\prime$ are obtained from $p$ and $p^\prime$ 
by interchanging the plus and minus components and changing the signs of the transverse components. 
In Eq.~(\ref{eq:particle-momenta}) we have introduced the, so-called, \lq\lq skewness\rq\rq\ $\xi$ 
which is defined as $\xi \, := \, (-\Delta^+)/(2\bar{p}^+)$. It parametrizes the relative momentum transfer 
between the proton and the $\overline{D^0}$ in the light-cone plus direction.

Considering $\ourprocess$ in the double-handbag mechanism means that the process amplitude 
can be split up into a nonperturbative part describing the soft hadronic $p \to \overline{D^0}$
and $\bar{p} \to D^0$ transitions and into a perturbative subprocess on the partonic level. 
In the latter only the minimal number of hadronic constituents which are necessary to convert the 
initial $p\bar{p}$ into the final $\overline{D^0}D^0$ pair take part actively. 
In our investigations we only take the valence Fock-state contributions into account. 
Furthermore, the proton is treated within a quark-diquark picture, where only scalar diquarks are considered.    
Thus it is the subprocess $S[ud]\overline{S[ud]} \,\to\, \bar{c}c$ which contributes on the partonic level.  
The virtuality of the intermediate gluon in the subprocess has to be high enough to produce the heavy $\bar{c}c$ pair. 
I.e., the heavy charm-quark mass sets a natural hard scale in our process, such that the process 
$S[ud]\overline{S[ud]} \,\to\, \bar{c}c$ can be treated by means of perturbative QCD. 
In order to argue that we can factorize the hard and the soft parts in this way we follow a similar line of argumentation as 
in Refs.~\cite{Goritschnig:2009sq, DFJK1}, where one poses physically motivated assumptions on the parton momenta.  
In our framework the $\ourprocess$ amplitude then reads: 
\begin{equation}
\begin{split}
 M_{\mu\nu} \,=\,
 & \frac{1}{4(\bar{p}^+)^2} \, \sum_{\lambda_1^\prime,\,\lambda_2^\prime} \,
 \int d\bar{x}_1 \, \int d\bar{x}_2 \, H_{\lambda_1^\prime,\,\lambda_2^\prime}(\bar{x}_1,\,\bar{x}_2) \,
 \frac{1}{\bar{x}_1-\xi} \, \frac{1}{\bar{x}_2-\xi} \, \\
 & \times \bar{v}(k_1^\prime,\,\lambda_1^\prime) \gamma^+ \bar{p}^+ \int \frac{d z_1^-}{2\pi} e^{\imath \bar{x}_1 \bar{p}^+ z_1^-}
 \langle \overline{D^0} :\, p^\prime \mid
 \Psi^c(-z_1^-/2) \Phi^{S[ud]}(+z_1^-/2)
 \mid p :\, p,\,\mu \rangle \\
 & \times \bar{p}^+  \int \frac{d z_2^+}{2\pi} e^{\imath \bar{x}_2 \bar{p}^+ z_2^+}
 \langle D^0 :\, q^\prime \mid
 \Phi^{S[ud]\,\dagger}(+z_2^+/2) \overline{\Psi}^c(-z_2^+/2)
 \mid \bar{p} :\, q,\,\nu \rangle
 \gamma^- u(k_2^\prime,\,\lambda_2^\prime) \,.
\label{eq:DHM-amplitude-simplified}
\end{split}
\end{equation}
$H_{\lambda_1^\prime,\,\lambda_2^\prime}(\bar{x}_1,\,\bar{x}_2)$ denotes the hard $S[ud]\overline{S[ud]} \,\to\, \bar{c}c$ amplitude, 
whereas the Fourier transforms of the hadronic matrix elements 
\begin{equation}
 \bar{p}^+ \, \int \frac{d z_1^-}{2\pi} e^{\imath \bar{x}_1 \bar{p}^+ z_1^-}
 \langle \overline{D^0} :\, p^\prime \mid \Psi^c(-z_1^-/2) \Phi^{S[ud]}(+z_1^-/2) \mid p :\, p,\,\mu \rangle
\label{eq:LC-p-to-Dbar-transition-matrix-element}
\end{equation}
and 
\begin{equation}
\bar{p}^+  \int \frac{d z_2^+}{2\pi} e^{\imath \bar{x}_2 \bar{p}^+ z_2^+}
 \langle D^0 :\, q^\prime \mid
 \Phi^{S[ud]\,\dagger}(+z_2^+/2) \overline{\Psi}^c(-z_2^+/2)
 \mid \bar{p} :\, q,\,\nu \rangle
\label{eq:LC-pbar-to-D-transition-matrix-element}
\end{equation}
incorporate the soft $p \,\to\, \overline{D^0}$ and $\bar{p} \,\to\, D^0$ transitions, respectively. 
Eq.~(\ref{eq:DHM-amplitude-simplified}) now is a convolution integral between the hard and the soft parts with respect to the 
average momentum fractions $\bar{x}_1 \,:=\, \bar{k}_1^+/\bar{p}^+$ and $\bar{x}_2 \,:=\, \bar{k}_2^-/\bar{p}^+$, 
where $\bar{k}_i \,:=\, (1/2) (k_i + k_i^\prime),\, i=1,2$. 

In the hard part of Eq.~(\ref{eq:DHM-amplitude-simplified}) the annihilation of the $S[ud]\overline{S[ud]}$ and 
the creation of the $\bar{c}c$ pair is separated by a lightlike distance (i.e., happens at the same light-cone time) and 
the active parton momenta can be approximated as being proportional to their parent hadron momenta. 
The soft hadronic transitions (\ref{eq:LC-p-to-Dbar-transition-matrix-element}) and (\ref{eq:LC-pbar-to-D-transition-matrix-element})
are hadronic matrix elements with respect to the incoming (anti)proton and the outgoing $\overline{D^0}$ ($D^0$) states of appropriate 
$c$-quark and $S[ud]$-diquark field operators. The action of the field operators in Eq.~(\ref{eq:LC-p-to-Dbar-transition-matrix-element}) is
that $\Phi^{S[ud]}(+z_1^-/2)$ emits an $S[ud]$ diquark from the proton into the partonic subprocess, whereas 
$\Psi^c(-z_1^-/2)$ reinserts a $\bar{c}$ quark into the remnants of the proton in order to give the final $\overline{D^0}$. 
The emission and the reinsertion (i.e. the arguments of the field operators) are again separated by a lightlike distance. 
An analog interpretation holds for the hadronic matrix element in Eq.~(\ref{eq:LC-pbar-to-D-transition-matrix-element}). 
It should be noted that as in Ref.~\cite{Goritschnig:2009sq} we have used appropriate projection techniques in deriving 
Eq.~(\ref{eq:DHM-amplitude-simplified}), such that only the, in our kinematical situation, dominant contributions of the field operators survive.

\section{The soft and hard contributions of $\ourprocess$}
\label{sec:The-soft-and-hard-contributions}

In order to model the soft $p \,\to\, \overline{D^0}$ and $\bar{p} \,\to\, D^0$ transitions, 
Eqs.~(\ref{eq:LC-p-to-Dbar-transition-matrix-element}) and (\ref{eq:LC-pbar-to-D-transition-matrix-element}) respectively, 
we will exploit the advantages of light-cone quantum field theory and 
derive an overlap representation in terms of hadronic LCWFs \cite{Diehl:2000xz}.   
In the following we will only discuss the $p \,\to\, \overline{D^0}$ transition, 
since one can proceed in complete analogy for the $\bar{p} \,\to\, D^0$ one. 
Within the light-cone quantum field theoretical framework we apply the Fourier representation of the field operators 
as well as the Fock-state decomposition of the hadron states in Eq.~(\ref{eq:LC-p-to-Dbar-transition-matrix-element}). 
Each Fock-state component, $\mid S[ud]u \,\rangle$ for the proton and $\mid \bar{c}u \,\rangle$ for the $\overline{D^0}$, 
comes with a LCWF as a \lq\lq prefactor\rq\rq. 
Furthermore, we only take into account LCWFs where we neglect partonic orbital angular momenta, 
such that the helicities of the partons add up to the parent hadron helicity. 
Thus we need only two LCWFs: one for the proton and one for the $\overline{D^0}$, 
which we call $\psi_p$ and $\psi_D$, respectively. 
Such hadronic LCWFs only depend on the relative parton momenta with respect to the parent hadron momentum, 
but not on the total hadron momentum. 

Before stating the result for the $\ourprocess$ amplitude after applying the overlap representation, 
we first take into account the fact that according to Refs.~\cite{Korner:1991zx, Kniehl:2005and2006} the D-meson wave function 
is strongly peaked around $x_0 \approx m_c/M$ with respect to its momentum fraction dependence. 
This behavior is also inherited by the LCWF overlap such that it notably weights momentum fraction close to the peak position 
in the hard scattering amplitude. It is therefore possible to replace the momentum fractions inside the hard-scattering amplitude by 
their peak position and to take the subprocess amplitude out of the convolution integral. 
Then the $\ourprocess$ amplitude takes on the simpler form
\begin{equation}
\begin{split}
 M_{\mu\nu} \,=\,
 & 2 {\mu\nu} \, H_{-\mu,\,-\nu}(x_0,\,x_0) \, \Big[
   \int d\bar{x} \,
   \frac{1}{\sqrt{\bar{x}^2-\xi^2}}
   \int \frac{d^2\bar{k}_{1\perp}}{16\pi^3} \\
 & \times \psi_D(\hat{x}^\prime(\bar{x}_1,\,\xi),\,\mathbf{\hat{k}}^\prime_\perp(\mathbf{\bar{k}}_{1\perp},\,\bar{x}_1,\,\xi)) \,
    \psi_p(\tilde{x}(\bar{x}_1,\,\xi),\,\mathbf{\tilde{k}}_\perp(\mathbf{\bar{k}}_{1\perp},\,\bar{x}_1,\,\xi)) \Big]^2 \,,
\label{eq:DHM-amplitude-peaking-approximation}
\end{split}
\end{equation}
with the hard subprocess amplitudes calculated within perturbative QCD
\begin{equation}
\begin{split}
& H_{++} \,=\, + \, 4\pi\alpha_s(x_0^2 s) \, F_s(x_0^2 s) \, \frac{4}{9} \frac{2M}{\sqrt{s}} \, \cos\theta \,,\quad
  H_{+-} \,=\, - \, 4\pi\alpha_s(x_0^2 s) \, F_s(x_0^2 s) \, \frac{4}{9} \, \sin\theta \,, 
\label{eq:hard-subprocess-amplitudes}
\end{split}
\end{equation}
$H_{--} \,=\, - H_{++}$ and $H_{-+} \,=\, H_{+-}$. 
We have added the color factor $4/9$ to the subprocess amplitudes, $\alpha_s$ is the strong coupling constant
and the phenomenological form factor $F_s$ takes care of the composite nature of the $S[ud]$ diquark at the $SgS$ vertex, 
cf. Refs.~\cite{Anselmino:1987vk, Anselmino:1987gu, Kroll:1993zx}.

\section{Model results}
\label{sec:Model-results}

In order to obtain model results for the $\ourprocess$ cross sections we have to specify the model LCWFs 
for the proton and the D meson which appear in the overlap representation.
We take the wave functions as have been used in Refs.~\cite{Goritschnig:2012vs, Kroll:1988cd}, 
\begin{equation}
 \psi_p(\tilde{x},\,\mathbf{\tilde{k}}_\perp) \,=\,
 N_p \, \tilde{x} \, e^{- a_p^2 \frac{\mathbf{\tilde{k}}_\perp^2}{\tilde{x}(1-\tilde{x})}}
\quad\text{and}\quad
 \psi_D(\hat{x}^\prime,\,\mathbf{\hat{k}}_\perp^\prime) \,=\,
 N_D \, e^{- a_D^2 \frac{\mathbf{\hat{k}}_\perp^{\prime 2}}{\hat{x}^\prime(1-\hat{x}^\prime)}} \,
  e^{- a_D^2 M^2 \frac{(\hat{x}^\prime-x_0)^2}{\hat{x}^\prime(1-\hat{x}^\prime)}}
\label{eq:LCWFs}
\end{equation}
for the proton and the $\overline{D^0}$, respectively. 
The arguments of the LCWFs are the relative momentum fractions $\tilde{x}$ and $\hat{x}^\prime$ and 
the intrinsic transverse momenta $\mathbf{\tilde{k}}_\perp$ and $\mathbf{\hat{k}}_\perp^\prime$ of 
the active constituents inside the proton and the $\overline{D^0}$, respectively, with respect to their parent hadron momenta. 
The $\overline{D^0}$ wave function has its peak around $x_0$ through its Gaussian mass exponential. 
The free parameters appearing in each wave function, namely the normalization constant $N_{p/D}$ 
and the transverse size parameter $a_{p/D}$, can be fixed by requirering a specific value for the root-mean-square 
of the intrinsic transverse momentum of the active constituent and the valence Fock-state probability or decay constant. 
I.e., the model parameters can thus be associated with physical parameters. 
Choosing $P_p = 0.5$ and $\langle \mathbf{k}_\perp^2 \rangle_p^{1/2} = 280\text{MeV}$ for the proton and 
$P_D = 0.9$ and $f_D = 206\text{MeV}$ for the D meson \cite{PDG} gives: 
$N_p = 61.8 \text{GeV}^{-2}$, $a_p = 1.1\text{GeV}^{-1}$ and $N_D = 55.2\text{GeV}^{-2}$, $a_D = 1.1\text{GeV}^{-1}$. 

On the right panel of Fig.~\ref{fig:overlap} we show the overlap integral with respect to the $\bar{x}$ and the 
$\mathbf{\bar{k}}_\perp$ integration which appears within the square brackets of Eq.~(\ref{eq:DHM-amplitude-peaking-approximation})
when using the wave functions (\ref{eq:LCWFs}) with the parameters as chosen above. 
We have plotted it as a function of $\cos\theta$ for different values of Mandelstam $s$, 
namely $s = 15,\,20$ and $30\text{GeV}^2$ corresponding to the dotted, dashed and solid lines, respectively. 
One observes that the magnitude of the overlap decreases with increasing CMS energies $s$ and, 
for all values of Mandelstam $s$, it shows an increase when going to small scattering angles. 

\begin{figure*}[h]
 \includegraphics[width=0.48\textwidth, height=0.25\textwidth]{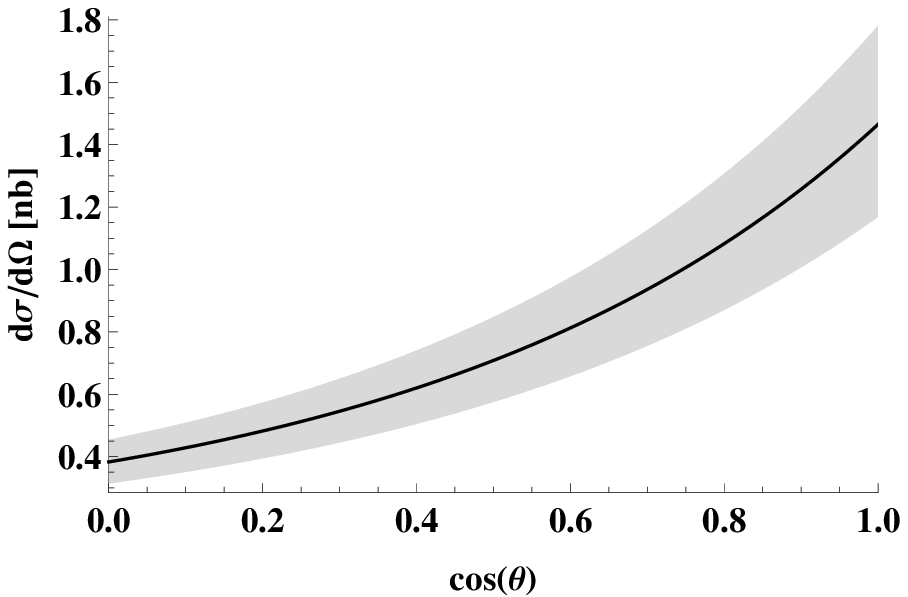}\hfill
 \includegraphics[width=0.48\textwidth, height=0.25\textwidth]{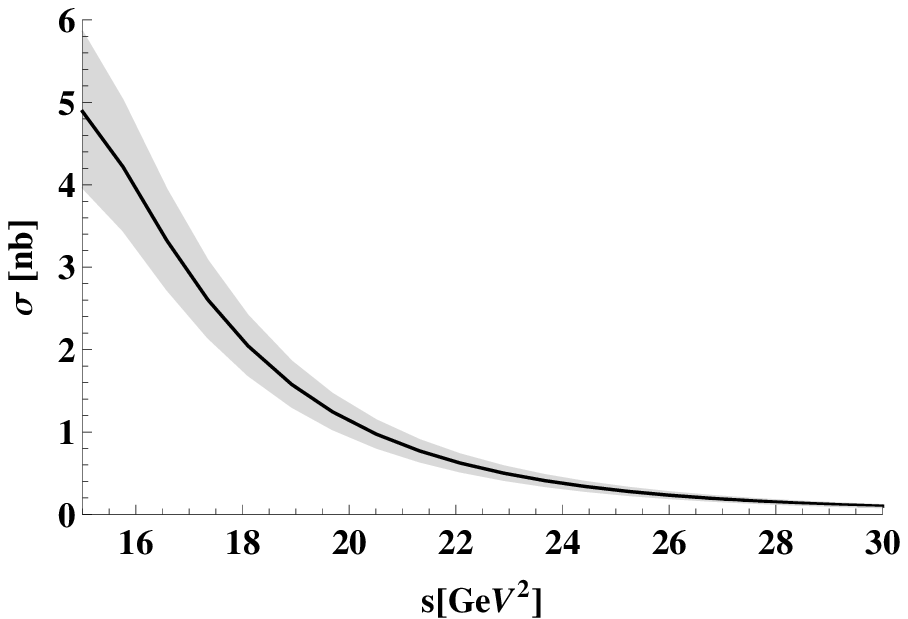}
\caption{{\it Left:} The differential cross section
$d\sigma_{p\bar{p}\rightarrow\overline{D^0}D^0}/d\Omega$ for
$s=15\,\text{GeV}^2$ versus $\cos\theta$. 
{\it Right:} The integrated $\ourprocess$ cross section as a function of $s$.}
\label{fig:cross_sections}
\end{figure*}

On the left-hand side of Fig.~\ref{fig:cross_sections} we present our predictions for the differential $\ourprocess$ cross section
$d\sigma_{p\bar{p}\rightarrow\overline{D^0}D^0}/d\Omega$ versus $\cos\theta$ for Mandelstam $s=15\,\text{GeV}^2$.
It is in the region of nb and it also shows the increase in magnitude for decreasing CMS scattering angle $\theta$. 
This property is inherited from the behavior of the overlap. 
On the right-hand side of Fig.~\ref{fig:cross_sections} we show our integrated cross-section predictions as a function of $s$. 
It is of the order of nb, which still is in the range $\bar{\text{P}}\text{ANDA}$ is able to measure. 
Our cross section predictions are in accordance with the results of Ref.~\cite{Kroll:1988cd}, where also a quark-diquark model has been used,
but are one order of magnitude smaller than the hadronic interaction-model calculations of Refs.~\cite{Khodjamirian:2011sp, Haidenbauer:2010nx}.

\section{Summary}
\label{sec:Summary}
We have investigated the process $\ourprocess$ within a double handbag approach where the process amplitude can be factorized into 
a hard subprocess amplitude on the constituent level and soft hadronic $p \,\to\, \overline{D^0}$ and $\bar{p} \,\to\, D^0$ matrix elements.  
We have treated the hard subprocess perturbatively and modelled the soft hadronic transitions as a LCWF overlap. 
In doing so we have obtained predictions for the differential and integrated $\ourprocess$ cross sections.

\ack
We would like to thank the organizing committee of the \lq\lq FAIRNESS 2013\rq\rq\ workshop for their efforts. 
ATG is supported by the Austrian science fund FWF under Grant No. J 3163-N16.
%  \medskip

\end{document}